\documentclass[two column, superscriptaddress, groupedaddress, amssymb, amsmath, nobibnotes, nofootinbib, aps, showpacs]{revtex4}
\usepackage[dvips]{graphicx}
\usepackage{lineno}
\usepackage{graphicx}

\date{\today}

\begin{document}

\title{Contrasting chemical pressure effect on the moment direction in the Kondo semiconductor CeT$_2$Al$_{10}$ (T = Ru,Os)}

\author{D. T. Adroja} 
\email{devashibhai.adroja@stfc.ac.uk}
\affiliation{ISIS Facility, Rutherford Appleton Laboratory, Chilton, Didcot, Oxon, OX11 0QX, United Kingdom} 
\affiliation{Highly Correlated Matter Research Group, Physics Department, University of Johannesburg, Auckland Park 2006, South Africa}
\author{A. D . Hillier}
\affiliation{ISIS Facility, Rutherford Appleton Laboratory, Chilton, Didcot, Oxon, OX11 0QX, United Kingdom} 
\author{C. Ritter}
\affiliation{Institute Laue- Langevin, BP 156, 6 Rue Jules Horowitz 38042,  Grenoble Cedex, France}
\author{A. Bhattacharyya}
\affiliation{ISIS Facility, Rutherford Appleton Laboratory, Chilton, Didcot, Oxon, OX11 0QX, United Kingdom} 
\affiliation{Highly Correlated Matter Research Group, Physics Department, University of Johannesburg, Auckland Park 2006, South Africa}
\author{D. D. Khalyavin} 
\affiliation{ISIS Facility, Rutherford Appleton Laboratory, Chilton, Didcot, Oxon, OX11 0QX, United Kingdom}
\author{A. M. Strydom} 
\affiliation{Highly Correlated Matter Research Group, Physics Department, University of Johannesburg, Auckland Park 2006, South Africa}
\affiliation{Max Planck Institute for Chemical Physics of Solids, N\"othnitzerstr. 40, 01187 Dresden, Germany}
\author{P. Peratheepan}
\affiliation{Highly Correlated Matter Research Group, Physics Department, University of Johannesburg, Auckland Park 2006, South Africa}
\affiliation{Department of Physics, Eastern University, Vantharumoolai, Chenkalady 30350, Sri Lanka}
\author{B. F\aa k}
\author{M. M. Koza}
\affiliation{Institute Laue- Langevin, BP 156, 6 Rue Jules Horowitz 38042,  Grenoble Cedex, France}
\author{J. Kawabata}
\author{Y. Yamada}
\author{Y. Okada}
\affiliation{Department of Quantum Matter, ADSM and IAMR, Hiroshima University, Higashi-Hiroshima 739-8530, Japan}
\author{Y. Muro}
\affiliation{Faculty of Engineering, Toyama Prefectural University, Toyama 939-8530, Japan}
\author{T. Takabatake}
\affiliation{Department of Quantum Matter, ADSM and IAMR, Hiroshima University, Higashi-Hiroshima 739-8530, Japan}
\author {J. W. Taylor}
\affiliation{ISIS Facility, Rutherford Appleton Laboratory, Chilton, Didcot, Oxon, OX11 0QX, United Kingdom} 

\begin{abstract}
The opening of a spin gap in the orthorhombic  compounds CeT$_2$Al$_{10}$ (T = Ru and Os) is followed by antiferromagnetic ordering at $T_N$ = 27 K and 28.5 K, respectively,  with a small ordered moment (0.29$-$0.34$\mu_B$) along the $c-$axis, which is not an easy axis of the crystal field (CEF). In order to investigate how the moment direction and the spin gap energy change with 10\% La doping in Ce$_{1-x}$La$_x$T$_2$Al$_{10}$ (T = Ru and Os) and also to understand the microscopic nature of the magnetic ground state, we here report on magnetic, transport, and thermal properties, neutron diffraction (ND) and inelastic neutron scattering (INS) investigations  on these compounds. Our INS study reveals the persistence of spin gaps of 7 meV and 10 meV in the 10\% La-doped T = Ru and Os compounds, respectively. More interestingly our ND study shows a very small ordered moment of 0.18 $\mu_B$ along the $b-$axis (moment direction changed compared with the undoped compound), in Ce$_{0.9}$La$_{0.1}$Ru$_2$Al$_{10}$, however a moment of 0.23 $\mu_B$ still along the $c-$axis in Ce$_{0.9}$La$_{0.1}$Os$_2$Al$_{10}$. This contrasting behavior can be explained by a different degree of hybridization in CeRu$_2$Al$_{10}$ and CeOs$_2$Al$_{10}$, being stronger in the latter than in the former. Muon spin rotation ($\mu$SR)  studies on Ce$_{1-x}$La$_x$Ru$_2$Al$_{10}$ ($x$ = 0, 0.3, 0.5 and 0.7), reveal the presence of coherent frequency oscillations indicating a long$-$range magnetically ordered ground state for $x$ = 0 to 0.5, but an almost temperature independent Kubo$-$Toyabe response between 45 mK and 4 K for $x$ = 0.7.  We will compare the results of the present investigations with those reported on the electron and hole$-$doping in CeT$_2$Al$_{10}$.
\end{abstract}

\maketitle
\section{Introduction}

In recent years, the magnetic and transport properties of Ce-based ternary compounds of type CeT$_2$Al$_{10}$ (T = Fe, Ru and Os), which crystalline in the orthorhombic structure (space group No 63 Cmcm)~\cite{Thiede}, have generated strong interest in both theoretical and experimental condensed matter physics~\cite{1,2,3,ddk,4,dta,5,6,7,8}. This interest arose due to the various ground states observed in this family of Ce-compounds. An unusually sharp phase transition near 27 K in the magnetic susceptibility of CeRu$_2$Al$_{10}$ has been attributed to a spin-dimer formation ~\cite{9,10}. The resistivity of CeRu$_2$Al$_{10}$ exhibits a sharp drop near 27 K resembling an insulator-metal transition~\cite{1}. A very similar phase transition, near 29 K, has been observed in CeOs$_2$Al$_{10}$ ~\cite{5,11}, but in this compound the susceptibility (along the $a-$axis) exhibits a broad maximum near 45 K in contrast to a sharp drop at the phase transition (27 K) in CeRu$_2$Al$_{10}$ ~\cite{5,13}. The broad maximum in the susceptibility and its strong anisotropic behavior in the paramagnetic state reveal the presence of strong hybridization between 4$f$ and conduction electrons as well as strong single ion anisotropy arising from the crystal field potential~\cite{kyu,fst}.  

\par

Further difference between the two systems appears in the resistivity of the ordered state; the resistivity of CeOs$_2$Al$_{10}$ displays a thermal activation-type temperature dependence below 15 K while the resistivity of CeRu$_2$Al$_{10}$ exhibits a metallic behavior below the phase transition down to 2 K. In spite of the comparable transition temperatures a fundamental contrast in the electronic disposition of the two compounds has been exposed in recent high-pressure studies ~\cite{5}. Under a hydrostatic pressure of 1.75 GPa the electrical resistivity of CeRu$_2$Al$_{10}$  changes in a way as closely to match the overall behavior of the temperature dependent resistivity of CeOs$_2$Al$_{10}$ in zero applied pressure. Applying pressure is a well recognized tuning method of the 4$f-$ band with respect to the Fermi energy (E$_F$) in narrow-band systems. The results obtained for the two iso-electronic compounds are therefore an indication that the center of gravity of the 4$f-$band is lying on opposite sides of the E$_F$: the 4$f-$ spectral weight in CeRu$_2$Al$_{10}$ is most likely close to but below the E$_F$ whereas in CeOs$_2$Al$_{10}$, the 4$f$ band is most likely above the E$_F$ due to the more extended nature of the 5$d$ band of Os compared to the 4$d$ band of Ru. The 3$d$ transition metal compound CeFe$_2$Al$_{10}$ exhibits Kondo insulating behavior with a transport$-$derived gap of 15 K~\cite{2}, while an NMR study reveals a much larger value of the gap, namely 110 K~\cite{14}. 

\par
Neutron diffraction studies of CeT$_2$Al$_{10}$ (T=Ru and Os) reveal a very small ordered moments, 0.34 $\mu_B$ and 0.29 $\mu_B$, respectively, along the $c-$axis, which is not the direction expected from the single ion crystal field (CEF) anisotropy ~\cite{ddk,4,hk}. As the single ion crystal field would prefer the moment along the $a-$axis in both compounds, this indicates that the moment direction in these compounds is governed by the anisotropic magnetic exchange and not by the CEF anisotropy. The inelastic neutron scattering (INS) study at 4.5 K on the polycrystalline samples  of CeT$_2$Al$_{10}$ (T=Ru, Os and Fe) reveals a clear sign of a spin-gap formation of 8 meV, 11 meV and 12 meV, respectively~\cite{15, 16}. These gaps are nearly temperature independent up to 24 K, but disappear suddenly  at 27 K, 39 K and 75 K respectively~\cite{jmmp}. Above these temperatures, the INS response becomes very broad, of quasi-elastic-type. Very recently inelastic neutron scattering investigation on single crystals of CeRu$_2$Al$_{10}$, CeOs$_2$Al$_{10}$ and CeFe$_2$Al$_{10}$ have been performed~\cite{15,16,18}. Well defined gapped spin waves are observed in CeRu$_2$Al$_{10}$ and CeOs$_2$Al$_{10}$ that can well be explained by anisotropic exchange interactions. Even in the paramagnetic state of CeFe$_2$Al$_{10}$ (no magnetic ordering observed down to 50 mK)  the neutron study reveals a dispersive gapped magnetic excitations having the same propagation vector $k$ = (1, 0, 0) as observed in CeRu$_2$Al$_{10}$, suggesting that these magnetic excitations in the Kondo insulating state have some connection to the spin wave observed in the magnetically ordered state of CeT$_2$Al$_{10}$ (T = Ru and Os) ~\cite{15,16}.   

\par

The effects of electron (Ir/Rh) and hole (Re) doping on the transition metal site in CeT$_2$Al$_{10}$ (T = Ru and Os)  have been investigated, through magnetization, resistivity, muon spin rotation ($\mu$SR), and neutron scattering (both elastic and inelastic) ~\cite{7, 8, 19}. These  studies show the general trend that the hybridization between 4$f-$electrons and conduction electrons increases with hole-doping, while the Ce-4$f$ electrons become more localized with electron-doping. On hole-doping the spin gap and the antiferromagnetic order with an  anomalous direction of the magnetic moment (i.e. moment either along $c-$axis or $b-$axis)  not governed by the single ion crystal field anisotropy (this prefers moment along $a-$axis) survive with small ordered state moments of 0.18$-$0.23$\mu_B$ ~\cite{4,19, 20}. In contrast to this, electron doping destabilizes the spin gap formation and the antiferromagnetic ordering becomes normal with moment directions along the $a-$axis e.g.  governed  by the single ion anisotropy, and larger values of the ordered state moment, $\approx$ 1 $\mu_B$~\cite{21}. 

\begin{figure}[t]
\vskip -0.6 cm
\centering
\includegraphics[width = 7 cm]{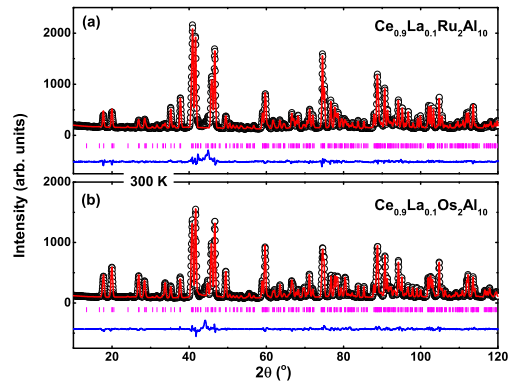}
\caption {(Color online) (a) and (b) show Rietveld refinement of the neutron powder diffraction pattern of Ce$_{0.9}$La$_{0.1}$T$_2$Al$_{10}$ (T = Ru and Os) at 300 K collected using the D2B diffractometer with wavelength $\lambda$ =1.594 \AA. The circle symbols (black) and solid line (red) represent the experimental and calculated intensities, respectively, and the line below (blue) is the difference between them. Tick marks indicate the positions of Bragg peaks in the Cmcm space group.}
\end{figure}

\par

Koyabasi {\it et. al.}~\cite{23} and Nishioka {\it et. al.}~\cite{5} suggested that the phase transition at $T_N$ is not due to simple RKKY interactions as it is not possible to describe the behavior of Ce$_{1-x}$Gd$_x$Ru$_2$Al$_{10}$ and Ce(Fe$_{1-x}$Ru$_x$)$_2$Al$_{10}$. Kondo {\it et. al.}~\cite{22} found that $T_N$ is suppressed by the application of a magnetic field as well as by La substitution on the Ce site in CeRu$_2$Al$_{10}$. Considering these interesting observations~\cite{pp,24} found with electron and hole doped CeRu$_2$Al$_{10}$ and CeOs$_2$Al$_{10}$, it is timely to investigate the effect of chemical pressure (La or Y doping) on the Ce site in CeT$_2$Al$_{10}$ (T = Ru and Os) using microscopic techniques such as neutron diffraction, inelastic neutron scattering and $\mu$SR measurements. We therefore present in this paper our results of such measurements  on La-substituted Ce$_{1-x}$La$_x$T$_2$Al$_{10}$ (T = Ru and Os) to shed light on the nature of the phase transition and the ground state of the Ce ion in these compounds. Our study is motivated by the highly unusual magnetic ordering in the Ru and Os compounds of the CeT$_2$Al$_{10}$ series, with questions about the enigmatic behavior first posed by Nishioka {\it et. al.}~\cite{5}. Moreover, in the special class of Kondo insulator materials, the Ru and Os compounds are to date the only cases where such a strongly hybridized and unstable $f-$shell condenses into a long-range magnetic ordered ground state. 

\par

The effect of La and Y substitution on the parent compound CeRu$_2$Al$_{10}$ has been investigated before by three groups by studying magnetic and transport properties~\cite{pp, 23, 24}. La (Y) has a bigger (smaller) ionic radius compared to Ce and hence La substitution expands the lattice corresponding to negative chemical pressure and Y  substitution contracts the lattice corresponding to that to positive chemical pressure. The results are interesting but unexpected~\cite{pp,23}. With increasing La concentration x in Ce$_{1-x}$La$_x$Ru$_2$Al$_{10}$, the transition temperature is progressively shifting to lower temperature and vanishes near the critical composition $x_c$ $\approx$ 0.7 ~\cite{24,yoy}. Surprisingly Y$-$substitution leads as well to a decrease of $T_N$ rather than to an increase as one would expect for positive chemical pressure, and disappears suddenly between $x$ = 0.4 and 0.5~\cite{yoy}. This behavior cannot be understood by a simple magnetic phase transition, suggesting that the change in the valence of Ce ion plays an important role in the mysterious phase transition ~\cite{25,yoy}. The high-field magnetization measurements on Ce$_{1-x}$La$_x$Ru$_2$Al$_{10}$ ($x$ = 0 and 0.25) performed by Kondo {\it et. al.} ~\cite{22} revealed that the long-range order disappears at a critical fields $H_c$ = 50 and 37 T for $x$ = 0 and 0.25, respectively.  

\par

From the single crystal susceptibility measurements Tanida {\it et. al.} have proposed that 10\% La doping in CeRu$_2$Al$_{10}$ changes the direction of the ordered state moment from the $c-$axis found in undoped system to the $b-$axis, the hard axis of magnetization. Application of a pressure of 0.3 GPa changes the moment back to the $c-$axis ~\cite{26}. However,magnetic susceptibility data  can only  give indirect  information on the ordered state moment direction and can be erroneous if anisotropic exchange interactions are dominating over the single ion crystal field anisotropy. They cannot give a direct measure of the value of the ordered state moment. The unanswered question remains what is happening to the spin gap formation and its energy scale as the direction of the ordered state moment of Ce$_{0.9}$La$_{0.1}$Ru$_2$Al$_{10}$  changes to the $b-$axis from the $c-$axis as in CeRu$_2$Al$_{10}$. In order to answer these questions we have carried out neutron diffraction and inelastic neutron scattering measurements on Ce$_{0.9}$La$_{0.1}$Ru$_2$Al$_{10}$. The neutron diffraction study provides direct information on the direction as on the magnitude of the ordered moment.  Inelastic neutron scattering gives direct information about the magnitude of the spin-gap energy, its temperature and wave-vector (Q) dependency. 

\par

In order to gain further information on the microscopic change in the magnetism we have performed muon spin rotation measurements on Ce$_{1-x}$La$_x$Ru$_2$Al$_{10}$ ($x$ = 0, 0.3, 0.5 and 0.7) alloys. $\mu$SR is an exceptionally sensitive microscopic probe of cooperative magnetic ordering phenomena and is thus ideally suited to our compounds where previous studied had alluded very small magnetic moment values.  Our neutron diffraction study reveals a long$-$range magnetically ordered ground state in both Ce$_{0.9}$La$_{0.1}$Ru$_2$Al$_{10}$ and Ce$_{0.9}$La$_{0.1}$Os$_2$Al$_{10}$ compounds. More interestingly the ordered Ce moment of 0.18 $\mu_B$ is along the $b-$axis in the former, but along the $c-$axis with a value of 0.23 $\mu_B$ for the latter.  Our INS study reveals the presence of a well defined spin gaps of 7 meV at 2 K in Ce$_{0.9}$La$_{0.1}$Ru$_2$Al$_{10}$ and 10 meV at 4.5 K with considerable reduced intensity in Ce$_{0.9}$La$_{0.1}$Os$_2$Al$_{10}$.

\begin{table}[b]
\begin{center}
\caption{Lattice parameters of Ce$_{1-x}$La$_{x}$T$_2$Al$_{10}$ (T = Ru, Os) for $x$ = 0 , 0.1 refined from the neutron diffraction data collected at 300 K in the orthorhombic Cmcm space group.}
\begin{tabular}{lccccccccccccccc}
\hline
Compounds && &&  a (\AA) && b (\AA)  && c (\AA) && V (\AA) &&\\ 
\hline
CeRu$_2$Al$_{10}$  &&  && 9.1246 && 10.2806 && 9.1878 && 861.9~\cite{ymuro}&& \\ 
Ce$_{0.9}$La$_{0.1}$Ru$_2$Al$_{10}$  &&  &&  9.1224  && 10.2749 && 9.1865  && 861.066 &&\\  
CeOs$_2$Al$_{10}$ &&  && 9.138 && 10.2662 && 9.1694 && 861.686~\cite{2}&&\\  
Ce$_{0.9}$La$_{0.1}$Os$_2$Al$_{10}$ &&  && 9.1412 && 10.2668 && 9.1898  && 862.474 && \\  
\hline
\end{tabular}
\end{center}
\end{table}

\begin{figure}[t]
\vskip 0.4 cm
\centering
\includegraphics[width = 9 cm]{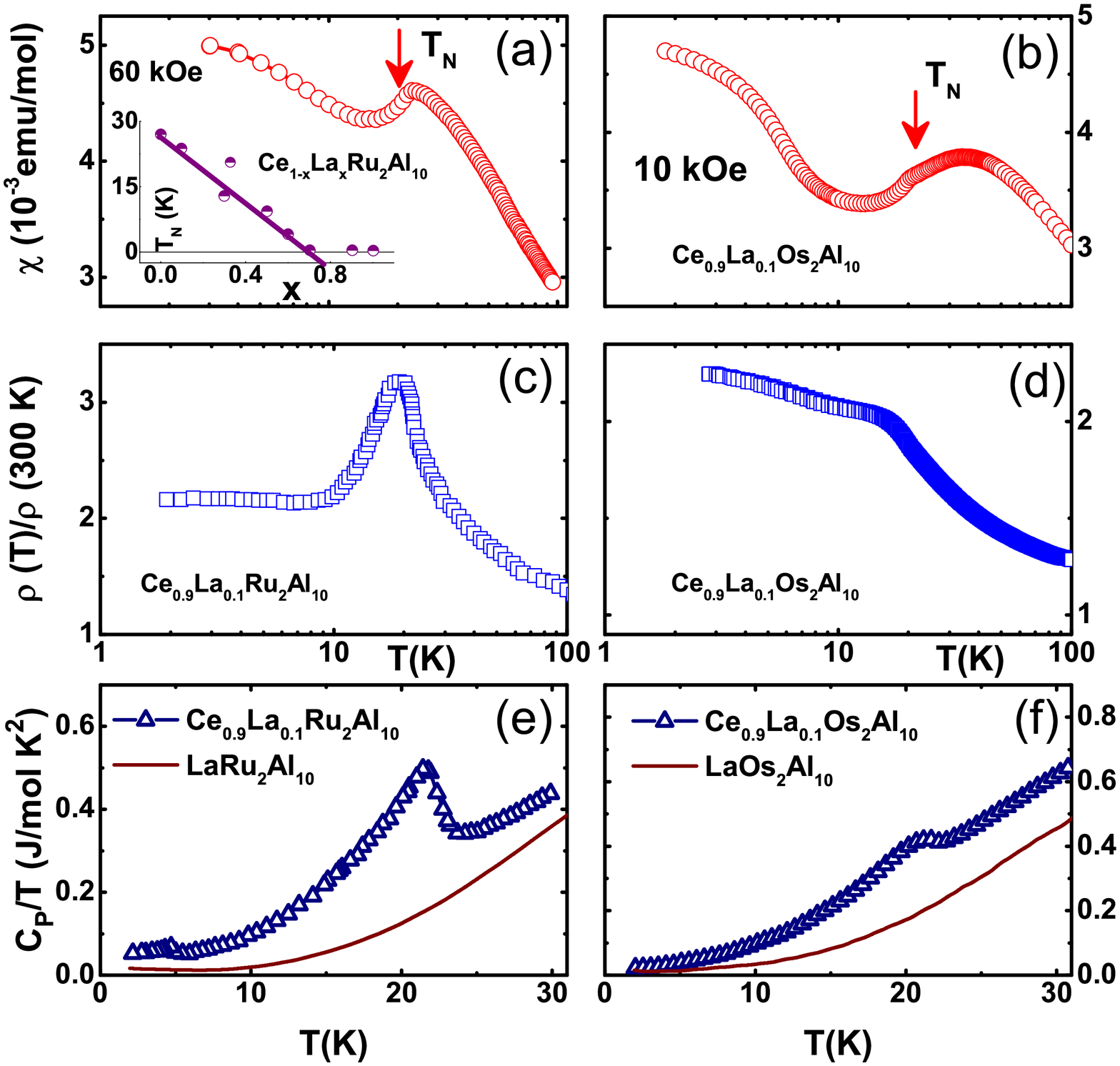}
\caption {(Color online) (a-b) Temperature dependence of dc magnetic susceptibility $\chi(T)$ Ce$_{1-x}$La$_x$T$_2$Al$_{10}$ with $x$ = 0.1 (T = Ru and Os). Inset shows the variation of $T_N$ with La composition for Ce$_{1-x}$La$_x$Ru$_2$Al$_{10}$~\cite{pp}. (c-d) Semilogarithmic plot of electrical resistivity vs temperature. (e-f) Temperature variation of specific heat $C_P$ divided by temperature for Ce$_{1-x}$La$_x$T$_2$Al$_{10}$ (T = Ru and Os) (open symbol) with non magnetic counterpart LaRu$_2$Al$_{10}$ and LaOs$_2$Al$_{10}$ (solid line).}
\end{figure}

\begin{figure}[t]
\vskip 0.4 cm
\centering
\includegraphics[width = 5 cm]{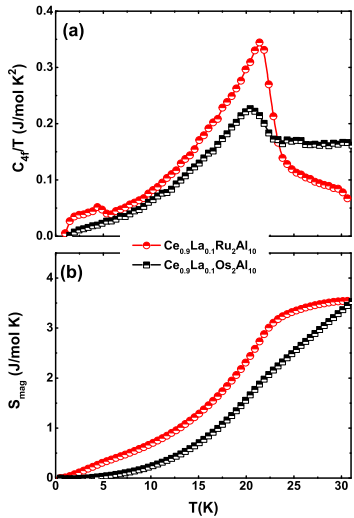}
\caption {(Color online) (a) Temperature variation of magnetic specific heat $C_{4f}$ for Ce$_{0.9}$La$_{0.1}$T$_2$Al$_{10}$ (T = Ru, Os). (b) Calculated magnetic entropy as a function of temperature.}
\end{figure}

\section{Experimental Details}

The polycrystalline samples of Ce$_{1-x}$La$_x$Ru$_2$Al$_{10}$ ($x$ = 0, 0.1, 0.3, 0.5 and 0.7), Ce$_{0.9}$La$_{0.1}$Os$_2$Al$_{10}$, LaRu$_2$Al$_{10}$ and LaOs$_2$Al$_{10}$ were prepared by argon arc melting of the stoichiometric constituents with the starting elements, Ce/La 99.9\% (purity), Ru/Os  99.9\%  and Al 99.999\%. The samples were annealed at 800 $^\circ$C for seven days in an evacuated quartz ampoule. The samples were characterized using powder X-ray diffraction or neutron diffraction (on D2B diffractometer at ILL, Grenoble) at 300 K and were found to be dominantly single-phase (see Fig. 1). Magnetic susceptibility measurements were made using a Magnetic Property Measurement System (MPMS) superconducting quantum interference device (SQUID) magnetometer (Quantum Design). Electrical resistivity by the four probe method and heat capacity by the relaxation method were performed in a Quantum Design Physical Properties Measurement System (PPMS). 

\par

The $\mu$SR experiments were carried out using the MUSR spectrometer in longitudinal geometry at the ISIS muon source, UK.  At the ISIS facility, a pulse of muons is produced every 20 ms and has a FWHM of $\approx$ 70 ns. These muons are implanted into the sample and decay with a half-life of 2.2 $\mu$s into a positron which is emitted preferentially in the direction of the muon spin axis. These positrons are detected and time stamped in the detectors which are positioned before, F, and after, B, the sample. The positron counts, N$_{F,B}$(t), have the functional form

\begin{equation}
N_{F,B}(t)=N_{F,B}(0)e^{-t/\tau_\mu}(1\pm G_z(t))          
\end{equation}

where G$_z$(t) is the longitudinal relaxation function. G$_z$(t) is determined using

\begin{equation}
G_z(t)=(N_F(t)-\alpha N_B(t))/(N_F(t)+ \alpha N_B(t))
\end{equation}

where $\alpha$ is a calibration constant which was determined at 35 K by applying a small transverse field ($\approx$ 20 Oe) and adjusting its value until the resulting damped cosine signal was oscillating around zero. The powdered samples were mounted onto a 99.995+\%  pure silver plate. 

\par

The low temperature neutron diffraction measurements at 1.5 K and 35 K on Ce$_{0.9}$La$_{0.1}$Ru$_2$Al$_{10}$ and Ce$_{0.9}$La$_{0.1}$Os$_2$Al$_{10}$ samples were performed using the high neutron flux D20 diffractometer at ILL, Grenoble, France using constant wavelengths of 1.3 \AA~or 2.41 \AA. The powder samples were mounted in a 10 mm diameter vanadium can, which was cooled down to 2 K using a standard He-4 cryostat. The program FULLPROF~\cite{ndr} was used for Rietveld refinements and group theoretical calculations were performed with the aid of the Sarah/ISOTROPY software ~\cite{27,iso}. 

\par

The inelastic neutron scattering measurements between 2 K and 35 K on Ce$_{0.9}$La$_{0.1}$Ru$_2$Al$_{10}$, Ce$_{0.9}$La$_{0.1}$Os$_2$Al$_{10}$, CeRu$_2$Al$_{10}$, LaRu$_2$Al$_{10}$ and LaOs$_2$Al$_{10}$ (15 g sample) were carried out using the MARI time-of-flight (TOF) chopper spectrometer at ISIS Facility, while on Ce$_{0.9}$La$_{0.1}$Ru$_2$Al$_{10}$ and LaRu$_2$Al$_{10}$ additional data were collected on the IN4 TOF chopper spectrometer at ILL, Grenoble, France. On MARI the samples were wrapped in a thin Al-foil and mounted inside a thin-walled cylindrical Al-can, which was cooled down to 4.5 K inside a top-loading closed-cycle-refrigerator (TCCR) with He-exchange gas around the samples. The measurements were performed with an incident neutron energy E$_i$ of 25 (20) meV, with an elastic resolution (at zero energy transfer) of 1.1 meV (0.8 meV) (FWHM). On IN4 the samples were wrapped in a thin Al-foil, which was cooled down to 2 K inside a standard He-4 cryostat with He-exchange gas around the samples. The measurements were performed with an incident neutron energy E$_i$ of 16.9 meV, with an elastic resolution (at zero energy transfer) of 1.1 meV  (FWHM).

\par

\begin{figure}[t]
\vskip 0.4 cm
\centering
\includegraphics[width = 9 cm]{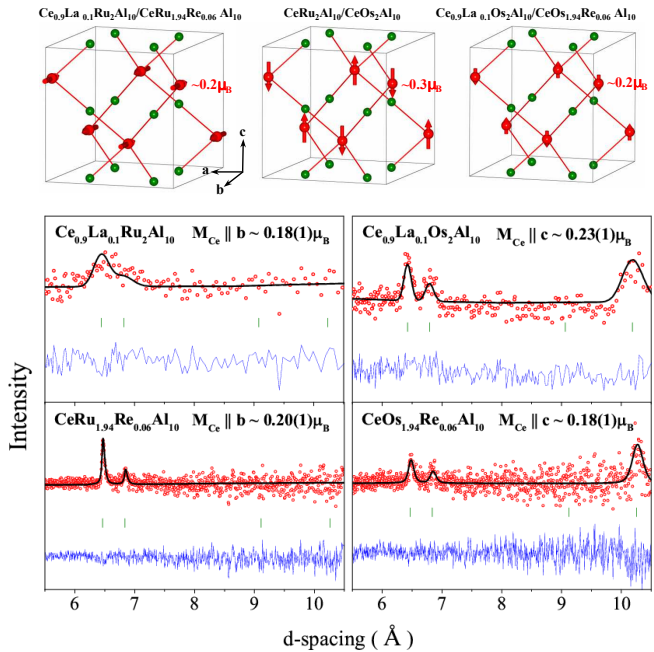}
\caption {(Color online) Rietveld refinements of the magnetic intensity of Ce$_{1-x}$La$_{x}$T$_2$Al$_{10}$ (T = Ru, Os) along with Ce(Ru/Os$_{1-x}$Re$_x$)$_2$Al$_{10}$ ($x$ = 0.03) obtained as a difference between the diffraction patterns collected at 1.5 K and 35  K. The circle symbols (red) and solid line represent the experimental and calculated intensities, respectively, and the line below (blue) is the difference between them. Tick marks indicate the positions of Bragg peaks for the magnetic scattering with the ${k}$= (1, 0, 0) propagation vector. The upper panel shows the magnetic structures of the La-doped (found in this paper), Ce(Ru/Os$_{1-x}$Re$_x$)$_2$Al$_{10}$ ($x$ = 0.03) and CeT$_2$Al$_{10}$ (T = Ru and Os) samples. For clarity, only Ce and Ru/Re atoms are shown (top).}
\end{figure}

\section{Results and discussions}
\subsection{Bulk properties}

To explain the overall bulk properties of the title compounds, we present here their magnetic susceptibility, electric resistivity, and specific heat data with emphasis on the magnetic phase transitions. Figs. 2 (a$-$f) show the temperature dependent magnetic susceptibility, electrical resistivity, and heat capacity of Ce$_{0.9}$La$_{0.1}$Ru$_2$Al$_{10}$ and Ce$_{0.9}$La$_{0.1}$Os$_2$Al$_{10}$.  The magnetic susceptibility $\chi(T)$ of Ce$_{0.9}$La$_{0.1}$Ru$_2$Al$_{10}$ exhibits a clear peak near 23.0 K, which is due to an antiferromagnetic ordering of Ce moments. On the other hand $\chi(T)$ of Ce$_{0.9}$La$_{0.1}$Os$_2$Al$_{10}$ exhibits a broad maxima near 40 K and a kink near 21.7 K. The former is due to an opening of the spin gap, while the latter is due to the onset of the antiferromagnetic ordering.  A very similar behavior of $\chi(T)$~\cite{7} with a broad maximum at  31 K (due to spin gap formation) above $T_N$= 23 K is observed for CeRu$_{1.96}$Re$_{0.06}$Al$_{10}$. The inverse magnetic susceptibilities of Ce$_{0.9}$La$_{0.1}$Ru$_2$Al$_{10}$ and Ce$_{0.9}$La$_{0.1}$Os$_2$Al$_{10}$ exhibit Curie-Weiss behavior between 50 K and 300 K. A linear least-squares fit to the data yields an effective magnetic moment $\mu_{eff}$ = 2.12 $\mu_B$ and a paramagnetic Curie temperature $\theta_p$ = $-$95 K for Ce$_{0.9}$La$_{0.1}$Ru$_2$Al$_{10}$ and $\mu_{eff}$ = 2.58 $\mu_B$ and $\theta_p$ = $-$175 K for Ce$_{0.9}$La$_{0.1}$Os$_2$Al$_{10}$. The value of the magnetic moment suggests that the Ce atoms are in their normal Ce$^{3+}$ valence state in both the compounds. The negative value of $\theta_p$ is in agreement with AFM ordering, a negative sign for the exchange interactions, and/or the presence of the Kondo effect. The larger negative value of $\theta_p$ of Ce$_{0.9}$La$_{0.1}$Os$_2$Al$_{10}$ compared to that of Ce$_{0.9}$La$_{0.1}$Ru$_2$Al$_{10}$ suggests a stronger hybridization in the former. The inset in the Fig. 2(a) shows the La composition dependence of $T_N$ of Ce$_{1-x}$La$_{x}$Ru$_2$Al$_{10}$, which reveals that $T_N$ decreases almost linearly and becomes zero near x$\ge$0.7~\cite{9,pp}. In this critical region of compositions the heat capacity exhibits a rise at low temperature suggesting the presence of non-Fermi-liquid (NFL) behavior close to a quantum critical point (QCP)~\cite{pp}. 

\par

Figs. 2(c) and (d) show the electrical resistivity $\rho(T)$ of Ce$_{0.9}$La$_{0.1}$Ru$_2$Al$_{10}$ and Ce$_{0.9}$La$_{0.1}$Os$_2$Al$_{10}$ samples, respectively. At high temperature $\rho(T)$ of both samples increases with decreasing temperature up to $T_N$. Then $\rho(T)$ of Ce$_{0.9}$La$_{0.1}$Ru$_2$Al$_{10}$ exhibits a peak near $T_N$ and remains metallic at low temperature, while for Ce$_{0.9}$La$_{0.1}$Os$_2$Al$_{10}$ the $\rho(T)$ shows a slope change at $T_N$, but still increases with further decrease in the temperature down to 2 K. This contrasting behavior of the low temperature resistivity is similar to that observed in the undoped compounds ~\cite{11, 13}. It is interesting to note that the resistivity of slightly electron-(8\% Ir) and hole-(2\% Re) doped CeOs$_2$Al$_{10}$ exhibits metallic behavior in all directions below $T_N$ ~\cite{19}.  

\par

Figs. 2 (e) and (f) show the heat capacity divided by temperature $C_P/T$ vs $T$ of  Ce$_{0.9}$La$_{0.1}$Ru$_2$Al$_{10}$ and Ce$_{0.9}$La$_{0.1}$Os$_2$Al$_{10}$ along with their respective nonmagnetic phonon reference compounds, LaRu$_2$Al$_{10}$ and LaOs$_2$Al$_{10}$. The heat capacity of Ce$_{0.9}$La$_{0.1}$Ru$_2$Al$_{10}$ exhibits a $\lambda-$type anomaly near $T_N$, while the anomaly is considerably suppressed in  Ce$_{0.9}$La$_{0.1}$Os$_2$Al$_{10}$. A rapid suppression of the heat capacity anomaly near $T_N$ was also observed both in CeRu$_{1.94}$Re$_{0.06}$Al$_{10}$ and   CeOs$_{1.96}$Re$_{0.04}$Al$_{10}$ ~\cite{8,19}. By fitting the high temperature heat capacity (above $T_N$) to $C_P/T$=$\gamma$+$\beta T^2$, we have estimated the Sommerfeld coefficient $\gamma$ = 0.125 J/mol-K$^2$ and $\beta$= 3.5$\times$10$^{-4}$  J/mol-K$^4$  for Ce$_{0.9}$La$_{0.1}$Ru$_2$Al$_{10}$ and $\gamma$ =0.121 J/mol-K$^2$ and $\beta$ = 5.6$\times$10$^{-4}$ J/mol-K$^4$    for Ce$_{0.9}$La$_{0.1}$Os$_2$Al$_{10}$. The observed values of $\gamma$ for both the compounds are smaller than those observed for undoped compounds, $\gamma$ = 0.2 J/mol-K$^2$ for CeRu$_2$Al$_{10}$ and $\gamma$ = 0.541 J/mol-K$^2$ for CeOs$_2$Al$_{10}$, indicating heavy fermion behavior in the undoped compounds. From the value of $\beta$ = (12$\pi^4$/5) ($nN_Ak_B/\Theta_D^3$), where $N_A$ and $k_B$ have the usual meaning, and $n$ = 13 is the number of atoms per f.u., we estimated the Debye temperature to $\Theta_D$ =416 K and 355 K for Ce$_{0.9}$La$_{0.1}$Ru$_2$Al$_{10}$  and Ce$_{0.9}$La$_{0.1}$Os$_2$Al$_{10}$ respectively.  Fig. 3 (a) shows the magnetic heat capacity variation with temperature for Ce$_{0.9}$La$_{0.1}$Ru$_2$Al$_{10}$  and Ce$_{0.9}$La$_{0.1}$Os$_2$Al$_{10}$. The weak anomaly for the former at temperature below 5 K may be attributed to impurity contribution. The value of magnetic entropy $S_{mag}$ [Fig. 3 (b)] at 30 K is 3.5 J/mol-K for  Ce$_{0.9}$La$_{0.1}$Ru$_2$Al$_{10}$ and 3.55 J/mol-K for  Ce$_{0.9}$La$_{0.1}$Os$_2$Al$_{10}$, which is much smaller than Rln(2) = 5.76 J/mol-K. The reduced magnetic entropy can be explained on the basis of the Kondo effect. We also estimated the gap in the spin wave by fitting $C_{mag(T)}$ data below $T_N$, and we find gaps 50 K for  Ce$_{0.9}$La$_{0.1}$Ru$_2$Al$_{10}$ and 60 K for  Ce$_{0.9}$La$_{0.1}$Os$_2$Al$_{10}$, whose values are approximately half of those reported for the undoped compounds~\cite{5}.

\begin{figure}[t]
\centering
\includegraphics[width = 9.1 cm]{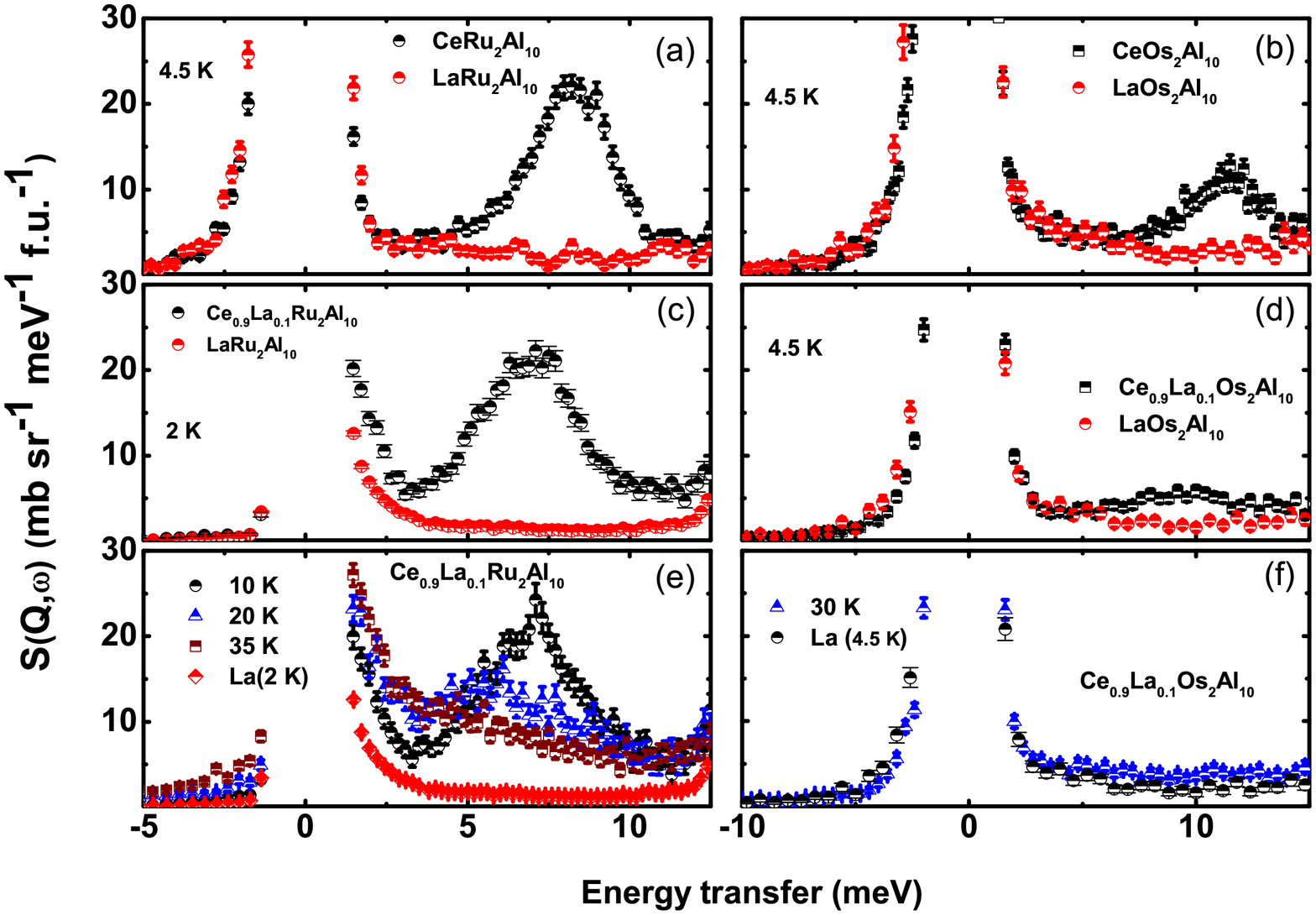}
\caption {(Color online) $Q$-integrated (0$\le$$Q$$\le$2.5~\AA) intensity versus energy transfer of (a-b) CeT$_2$Al$_{10}$ (T = Ru and Os) measured on the MARI spectrometer (c-f) Ce$_{0.9}$La$_{0.1}$Ru$_2$Al$_{10}$ and Ce$_{0.9}$La$_{0.1}$Os$_2$Al$_{10}$ measured on the IN4 spectrometer along with the nonmagnetic phonon reference compounds LaT$_2$Al$_{10}$ (T = Ru and Os) (red half filled circles), measured with respective incident energy of $E_i$ = 20 meV. For IN4 we have made cuts in scattering angle between 13$^\circ$ to 43$^\circ$.}
\end{figure}

\section{Neutron diffraction}

Figs. 1 (a) and (b) show the neutron diffraction patterns for Ce$_{0.9}$La$_{0.1}$Ru$_2$Al$_{10}$ and Ce$_{0.9}$La$_{0.1}$Os$_2$Al$_{10}$, collected at 300 K on the D2B instrument in the high resolution mode which both are consistent with the Cmcm symmetry and can be satisfactorily fitted with the structural model proposed earlier~\cite{Thiede}. The structural parameters for both samples are listed in Table I. A comparison of the lattice parameters of the doped compounds with the parent ones shows that 10\% La doping in CeRu$_2$Al$_{10}$ results in a volume contraction of 0.001, while 10\% La doping in CeOs$_2$Al$_{10}$ increases the volume by 0.001. This can be compared to the volume contraction of 0.10\% observed in 3\% Re-doped (or hole doped) CeRu$_2$Al$_{10}$, where the ordered state moment of 0.20 $\mu_B$ is along the $b-$axis. It is generally agreed that hole-doping in both CeRu$_2$Al$_{10}$ and CeOs$_2$Al$_{10}$ increases the hybridization~\cite{7, 20}. 

\par

To investigate the magnetic structure, we carried out neutron diffraction measurements on the D20 instrument at 1.5 K and 35 K (12 hours each temperature) on Ce$_{0.9}$La$_{0.1}$Ru$_2$Al$_{10}$ and Ce$_{0.9}$La$_{0.1}$Os$_2$Al$_{10}$. At 1.5 K we observed very weak magnetic Bragg peaks at scattering angles away from the nuclear Bragg peaks, which confirmed the long$-$range antiferromagnetic ordering of the Ce moment at 1.5 K in both compounds. To see the magnetic Bragg peaks clearly, we plotted the difference between the 1.5 K and 35 K data for both compounds, which are shown in Fig. 4, along with their 3\% Re doped counterparts (see Refs ~\cite{7, 20} for details of the neutron diffraction experiment of these materials). The important observation is the qualitatively similar diffraction patterns between the La and Re doped compositions. The absence of the magnetic (0, 1, 0) peak near 10.3 \AA~in both Ce$_{0.9}$La$_{0.1}$Ru$_2$Al$_{10}$ and CeRu$_{1.94}$Re$_{0.06}$Al$_{10}$ indicates the ordered moments to be along the $b-$axis, in contrast with the parent CeRu$_2$Al$_{10}$ compound as well as with the Ce(Ru$_{1−x}$Fe$_x$)$_2$Al$_{10}$ series, where the moments were found to be along the $c-$axis ~\cite{7}. The conclusion about the $b-$axis moment direction comes directly from the fact that the magnetic neutron diffraction intensity is proportional to the square of the ordered moment component perpendicular to the scattering vector. On the other hand in Ce$_{0.9}$La$_{0.1}$Os$_2$Al$_{10}$ several magnetic Bragg peaks were observed including the (0, 1, 0) one and the relative intensities of these peaks are very similar to those observed in CeRu$_2$Al$_{10}$, indicating that the moments are probably along the $c-$axis. In spite of the different moment directions, all the observed magnetic Bragg peaks in Ce$_{0.9}$La$_{0.1}$Ru$_2$Al$_{10}$ and Ce$_{0.9}$La$_{0.1}$Os$_2$Al$_{10}$ can be indexed based on the same propagation vector $k$ = (1, 0, 0), which is identical to that found in pure CeRu$_2$Al$_{10}$ ~\cite{3}. It is to be noted that the propagation vector $k$ = (0, 1, 0) proposed for CeOs$_2$Al$_{10}$ is equivalent to $k$ = (1, 0, 0) as they are both related by the allowed ($h$+$k$ even) reciprocal translation G = (-1, 1, 0).

\par

In the qualitative refinement of the magnetic structures, we employed a method whereby combinations of axial vectors localized on the 4$c$(Ce) site (as Ce is only the magnetic atom) and transforming as basis functions of the irreducible representations of the wave vector group are systematically tested. The symmetry analysis yields that the reducible magnetic representation is decomposed into six one-dimensional representations, labeled Y$^+_i$ ($i$ = 2,3,4) and Y$^-_i$ ($i$ = 1,2,3). The Y$^+_i$ representations result in a ferromagnetic (FM) alignment of the Ce moments within the primitive unit cell, along different crystallographic directions. On the other hand, Y$^-_i$ transform Ce moments which are AFM coupled within the primitive unit cell. We refined the difference data (1.5 K$-$35 K) using the three AFM structures (with the moments along a$-$, b$-$ and c$-$axes) given by Y$^-_i$ representations. The best fit to the data was obtained with a Ce ordered state moment of 0.18(2)$\mu_B$ AFM coupled along the $b-$axis for Ce$_{0.9}$La$_{0.1}$Ru$_2$Al$_{10}$ [see Fig. 4, where the magnetic unit cells are shown], while an ordered moment of 0.23(1) $\mu_B$ along the $c-$axis was obtained for Ce$_{0.9}$La$_{0.1}$Os$_2$Al$_{10}$. 

\par

It is interesting to compare these values of the ordered state moments with those found in the undoped systems, 0.34(2) $\mu_B$ for CeRu$_2$Al$_{10}$ and 0.29(2) $\mu_B$ for CeOs$_2$Al$_{10}$ both along the $c-$axis ~\cite{7,8}. This comparison shows a reduction of the ordered moment in 10\% La doped systems, which might be due to a change in the Ce valence with La-doping as proposed in Ref.~\cite{25}. The different directions of the ordered state moment in 10\% La doped CeRu$_2$Al$_{10}$ and CeOs$_2$Al$_{10}$ is a surprising observation and can be only explained by assuming a different degree of hybridization (weak in the former material). This means that  10\% La-doped CeOs$_2$Al$_{10}$ sees anisotropic exchange interactions which are still similar to those in the undoped compound preserving hence the moment direction. It should be possible to change the moment direction as well in La$-$doped CeOs$_2$Al$_{10}$ to the $b-$axis by increasing the La content up to 20 or 30\% in CeOs$_2$Al$_{10}$. However, the absolute value of the moment would certainly reduce further with increased doping making its experimental detection difficult. The question remains why$-$despite having a smaller hybridization the ordered moment in Ce$_{0.9}$La$_{0.1}$Ru$_2$Al$_{10}$ is smaller in comparison to that found in Ce$_{0.9}$La$_{0.1}$Os$_2$Al$_{10}$.  A possible explanation could be that the Kondo effect along the $b-$axis is stronger than along the $c-$axis, which might screen the moment value.

\section{Inelastic neutron scattering study}

With the dramatic and contrasting changes observed in the moment direction and its absolute value in 10\% La-doped CeRu$_2$Al$_{10}$ and CeOs$_2$Al$_{10}$ it would be interesting to investigate directly the spin gap formation in these compounds using inelastic neutron scattering. Furthermore, Kawabata {\it et. al.}~\cite{19} reported that the suppression of $T_N$ is well correlated with the gap energy $\Delta$ as a function of electron-(Ir) and hole-(Re) doping and they conclude that the presence of the hybridization gap is indispensable for the AFM order at unusually high $T_N$ in CeOs$_2$Al$_{10}$. Thus the information on the spin gap energy scale in Ce$_{0.9}$La$_{0.1}$Ru$_2$Al$_{10}$ and Ce$_{0.9}$La$_{0.1}$Os$_2$Al$_{10}$ is very important. Therefore, we briefly report the INS spectra, which give direct information of the spin gap energy, below and above $T_N$ of Ce$_{0.9}$La$_{0.1}$Ru$_2$Al$_{10}$ and Ce$_{0.9}$La$_{0.1}$Os$_2$Al$_{10}$ and also of CeRu$_2$Al$_{10}$ and CeOs$_2$Al$_{10}$ for comparison in this section. A detailed report on the inelastic neutron scattering investigations on CeT$_2$Al$_{10}$ (T = Fe, Ru and Os) compounds can be found in Ref. ~\cite{2}. 

\par

Fig. 5 displays the inelastic neutron scattering  spectra of 10\% La-doped compounds with that of undoped compounds at two temperatures at low$-$Q measured on the MARI and IN4 spectrometers. There is a clear magnetic excitation centered around 7 meV and 10 meV in Ce$_{0.9}$La$_{0.1}$Ru$_2$Al$_{10}$ and Ce$_{0.9}$La$_{0.1}$Os$_2$Al$_{10}$, respectively  which may be compared to the 8 meV  and 11 meV excitations found in the parent compounds CeRu$_2$Al$_{10}$ and CeOs$_2$Al$_{10}$, respectively~\cite{2,7,8}. The value of the peak position can be taken as a measure of the spin gap energy in these compounds. These results show that in both 10\% La doped systems, despite the moment direction being different and the ordered state moments being reduced, a very small change in the spin gap energy scale is observed. On the other hand the intensity of the spin gap does not change much in  Ce$_{0.9}$La$_{0.1}$Ru$_2$Al$_{10}$, while a dramatic reduction in the intensity of the spin gap is observed in Ce$_{0.9}$La$_{0.1}$Os$_2$Al$_{10}$. An interesting observation is the existence of a clear low energy response at low Q in Ce$_{0.9}$La$_{0.1}$Ru$_2$Al$_{10}$ at 2 K. This signal is not observed in the non-magnetic phonon reference compound LaRu$_2$Al$_{10}$, which confirms its magnetic origin. Further the high resolution INS study on CeRu$_2$Al$_{10}$ by J. Robert {\it et. al.}~\cite{16} did not reveal any clear sign of the low energy or quasi-elastic excitation at 11 K. This reveals that the low energy excitation in Ce$_{0.9}$La$_{0.1}$Ru$_2$Al$_{10}$ has some relation with changes in the moment direction from the $c-$axis to the $b-$axis. It is an open question whether this could be the zero frequency mode observed in several magnetically ordered heavy fermion systems~\cite{bk1, nhv}. To understand this, spin wave measurements on single crystals of  Ce$_{0.9}$La$_{0.1}$Ru$_2$Al$_{10}$ are highly desirable. On the other hand within the resolution of the MARI experiment we could not see any clear sign of a low energy excitation at 4.5 K in Ce$_{0.9}$La$_{0.1}$Os$_2$Al$_{10}$; this excitation was also absent in CeOs$_2$Al$_{10}$ from the high resolution INS study~\cite{2}.  

\par

Now we discuss the temperature dependence of the spin gap excitation. For Ce$_{0.9}$La$_{0.1}$Ru$_2$Al$_{10}$ with increasing temperature to 10 K no dramatic changes were observed in the spectra, but at 20 K the spin gap energy decreases to 5.5 meV and its width increases to 1.26 meV. Further increase in the temperature to 25 K (same response at 30 K) the inelastic response continues to broaden. The data show two components, a low energy/quasi-elastic component with narrow linewidth and a second, distinctly broader component. The Q-dependent integrated intensity of Ce$_{0.9}$La$_{0.1}$Ru$_2$Al$_{10}$ between 5 and 8 meV at 2 K nearly follows the Ce$^{3+}$ magnetic form factor squared (F$^2$(Q), figure not shown), very similar to that observed in pure CeRu$_2$Al$_{10}$~\cite{4}.  The observed single ion type response of Ce$_{0.9}$La$_{0.1}$Ru$_2$Al$_{10}$ in the magnetic ordered state could be due to the fact that the observed spin wave scattering intensity in CeRu$_2$Al$_{10}$ single crystal is stronger near the zone boundary and considering the presence of powder averaging effect could give single-ion type behavior. It is to be noted that the single-ion type response is also observed in the inelastic response of the spin gap system CeRu$_4$Sb$_{12}$ (no magnetic ordering down to 2 K), which does not exhibit any long-range magnetic ordering down to 50 mK~\cite{29}. On the other hand, the deviation from a single-ion response is observed in the spin gap system CeFe$_4$Sb$_{12}$, where it was proposed that the intersite interactions between Ce and Fe are playing an important role ~\cite{30}. As the spin gaps in CeOs$_2$Al$_{10}$ and CeRu$_2$Al$_{10}$ open up below (or just above) the magnetic ordering temperature one would expect that the spin gap energy and its intensity would be strongly Q$-$dependent, but this is not the case. Furthermore, with increasing temperature to 30 K the response of Ce$_{0.9}$La$_{0.1}$Os$_2$Al$_{10}$ also becomes very broad and the intensity decreases considerably [Fig. 5 (f)] compared with 4.5 K.

\begin{figure}[t]
\centering
\includegraphics[width = 7 cm]{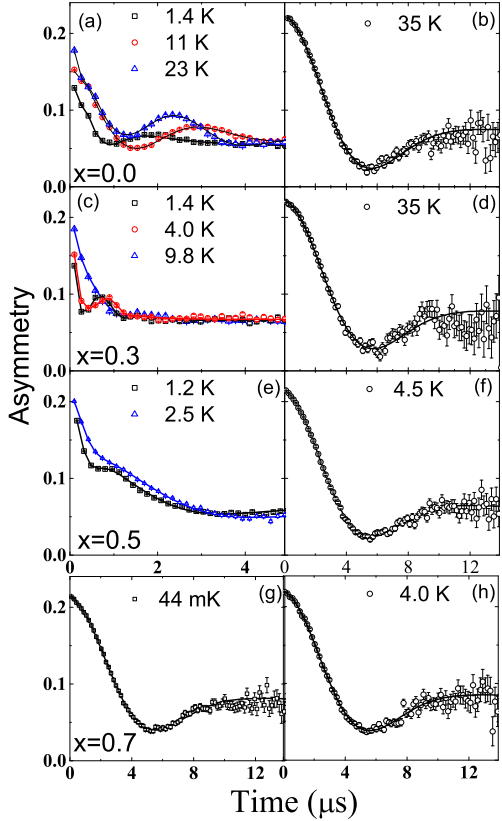}
\caption {(Color online) The time evolution of the muon spin relaxation in Ce$_{1-x}$La$_x$Ru$_2$Al$_{10}$ for various temperatures (above and below $T_N$) in zero field. The solid line is a least-squares fit [using Eq. (3) (above $T_N$) and Eq. (4) (below $T_N$)] to the data as described in the text.}
\end{figure}

\begin{figure}[t]
\vskip 0.4 cm
\centering
\includegraphics[width = 9 cm]{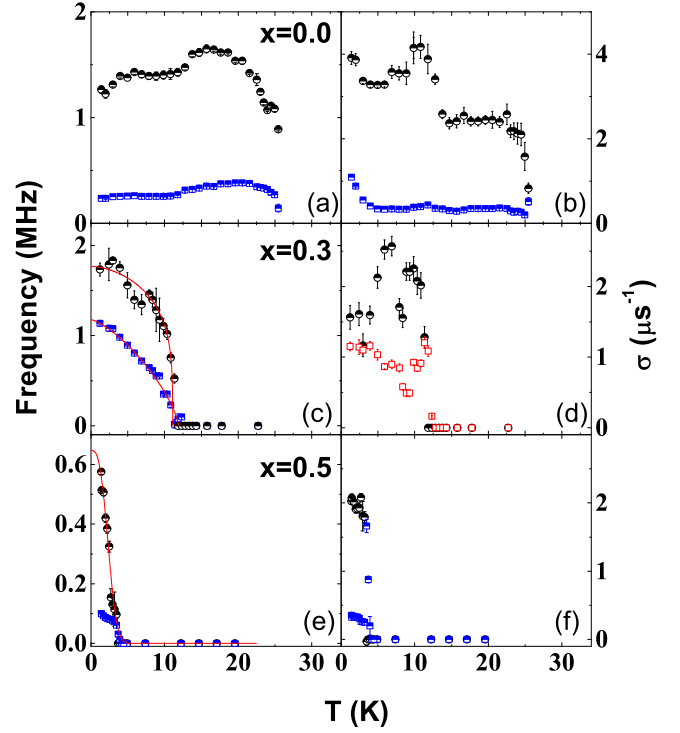}
\caption {(Color online) The temperature dependence of (a, c, e) muon precession frequency/internal field at the muon site in Ce$_{1-x}$La$_x$Ru$_2$Al$_{10}$. (b, d, f) the depolarization rate $\sigma$. The solid line in (c, e) is fit to the data using Eq. (4) (see text).}
\end{figure}

\section{Muon Spin relaxation}

Figs. 6(a$-$f) shows the zero-field (ZF) $\mu$SR spectra at various temperatures of Ce$_{1-x}$La$_x$Ru$_2$Al$_{10}$ ($x$ = 0, 0.3, 0.5 and 0.7). The left hand side figures show the spectra at low temperatures, while the right hand side figures show the spectra at high temperature. It is interesting to see the dramatic change in the time-evolution of the $\mu$SR spectra with temperature for all compositions, except for $x$ = 0.7. At 35 K (or 4.5 K)  we observe  a strong damping at shorter time, and the recovery at longer times for all compositions, which is a typical muon response to nuclear moment, known as the Kubo$-$Toyabe~\cite{31}, arising due to a static distribution of the nuclear dipole moment. Here it arises from the La (stable isotope $^{138}$La, I=5, 0.09\% abundance and $^{139}$La, I=7/2, 99.91\% abundance),  Ru (stable isotope $^{99}$Ru, I=5/2, 12.76\% abundance, $^{101}$Ru , I=5/2, 17.06\% abundance)  and Al (I=5/2) nuclear moment contributions (I = 0 for Ce, i.e. zero contribution). Above the magnetic ordering temperature the $\mu$SR spectra of all compounds ($x$ = 0 to 0.7) can all be described by the following equation (see Fig. 6, right hand figures): 

\begin{equation}
G_z (t)=A_0 \big[\frac{1}{3}  + \frac{2}{3}  (1-(\sigma t)^2 )e^{(-\sigma t)^2/2}\big]e^{-\lambda t}+A_{bg}  
\end{equation}

where $A_0$ is the initial asymmetry, $\sigma/\gamma_\mu$ is the local field distribution, $\gamma_\mu$=13.55 MHz/T is the gyromagnetic ratio of the muon, $\lambda$ is the electronic relaxation rate arising from electronic moments and $A_{bg}$ is a constant background. It is assumed that the electronic moments give an entirely independent muon spin relaxation channel in real time. The value of $\sigma$ was found to be 0.32(3) $\mu$s$^{-1}$ for all compositions from fitting the spectra at high temperature (35 K or 4.5/4 K)  to Eq. (3), which suggests that muon stopping sites are the same for all compositions. From a simple electrostatic potential calculations of CeRu$_2$Al$_{10}$ Kambe {\it et. al.}~\cite{32} have proposed muon stopping site at 4$a$ (0,0,0), while Khalyavin {\it et. al.}~\cite{ddk} proposed muon stopping site at 4$c$ (0.5,0,0.25) (using electrostatic calculation of CeOs$_2$Al$_{10}$). Further the Density Functional Theoretical (DFT) calculation of muon stopping sites in Ce(Ru$_{1-x}$Rh$_x$)$_2$Al$_{10}$~\cite{33} supports both muon sites equally possible as they are located at local minimum positions of the electrostatic potential. Further the DFT calculation revealed that the muon stopping site (or position) estimated from the potential calculation do not change much with Rh-doping~\cite{34} as there are no big changes in the potential around the suggested muon sites. 

\par

Now we discuss the muon spectra at low temperatures. As shown in the left side of Fig. 6, $\mu$SR spectra of $x$ = 0, 0.3 and 0.5 compounds exhibit a clear sign of coherent frequency oscillations confirming the long range magnetic ordering of the Ce moment. On the other hand $\mu$SR spectra of $x$ = 0.7 at 44 mK reveals the same behavior as that observed at 4.0 K, indicates the non-magnetic (or paramagnetic) ground state. The result is in agreement with the proposed phase diagram $T_N$ vs $x$ (La concentration), reveals $T_N \approx$0 K for $x\ge$ 0.6 [see the inset in Fig. 2 (a)]. For the other compositions, the spectra below $T_N$ are best described by two oscillatory terms and an exponential decay, as given by the following equation

\begin{equation}
G_z (t)=\sum_{i=1}^{2}A_i cos(\omega_i t+\phi)e^{-(\sigma t)^2/2}+A_3 e^{-\lambda t}+A_{bg} 
\end{equation}

where $\omega$=$\gamma_\mu$$H_{int}$ is the muon precession frequency ($H_{int}$ is the internal field at the muon site) and $\phi$ is the phase. In Fig. 7 (left) we have plotted the internal field (or muon precession frequency) at the muon site as a function of temperature. This shows that the two internal fields (or frequencies) appear just below 27 K in $x$ = 0, 12.5 K in $x$ = 0.3 and 4.2 K in $x$ = 0.5, showing a clear onset of bulk long-range magnetic order in agreement with $T_N$ proposed in the inset of Fig. 2 (a). It is interesting to note that even though the heat capacity of $x$ = 0.3 and 0.5 exhibits a very broad $\lambda-$ type anomaly near $T_N$ ~\cite{35}, the $\mu$SR shows a typical second order phase transition that can be explained by mean field behavior. Further, the associated internal fields are found to be very small in agreement with a small ordered magnetic moment of the Ce$^{3+}$ ion observed in the neutron diffraction for $x$ = 0 and 0.1. The observed two values of internal fields can be explained on the basis of the dipole field calculation. It was found that Kambe's suggested positions correspond to the 4$a$ sites which had the lower field and Khalyavin's suggested positions correspond to the 4$c$ sites which had the higher field. Further the temperature dependence of $\sigma$ and the differences between the two values of $\sigma$ also exhibit very similar behavior to that of the internal fields. The value of $\lambda$ was bound to be nearly temperature independent for all values of $x$.

\par
Now examining the temperature dependence of the internal fields, we can see that there is a dip in the internal field (see Fig. 7 top left), which occurs around 15 K. Moreover, below 15 K the first and the second component of the depolarization rates also increase (Fig. 7 right). In principle this could originate from various phenomena related to a change in distribution of internal fields, but a structural transition is a likely candidate in view of the structural instability reported on this system~\cite{1}. To find out the value of critical exponents and hence get more information on the nature of the magnetic transition, the temperature dependence of the internal field was fitted~\cite{ab}:

\begin{equation}
H_{int}(T)= H_0\left(1-\left(\frac{T}{T_N}\right)^{\alpha}\right)^{\beta}
\end{equation}

Fitting the temperature dependent internal field of $x$ = 0.3 to Eq. (5) we obtained the value of the parameters for higher [and for lower] internal field: $T_N$ =12.9(3) (same for both fields), $H_0$=16.5(3) Oe [11.8(7) Oe], $\alpha$ = 1.65(6) [1.42(4)] and $\beta$ = 0.81(3) [0.89(2)]. Fit for $x$ = 0.5 data fitting higher field we obtained $T_N$ = 4.06(9), $H_0$ = 9.6(20) Oe, $\alpha$ = 0.99(30) and $\beta$ = 1.2(12). It is to be noted that due to limited temperature range the fit to lower field did not converge. It is interesting to compare these values of the exponents with $\alpha$ = 1.47(2) and $\beta$ = 0.96 observed in  CeRu$_{1.94}$Re$_{0.06}$Al$_{10}$~\cite{7}.  The larger values of beta compared to 0.5, expected from the mean field theory suggests that magnetic interactions are complex in nature.

\par

Now we compare the results of our neutron diffraction and $\mu$SR of Ce$_{0.9}$La$_{0.1}$Ru$_2$Al$_{10}$ with that of the hole doped systems, Ce(Ru$_{0.97}$Re$_{0.03}$)$_2$Al$_{10}$ and Ce(Os$_{0.97}$Re$_{0.03}$)$_2$Al$_{10}$. The neutron diffraction study of Ce(Ru$_{0.97}$Re$_{0.03}$)$_2$Al$_{10}$ shows that the compound orders antiferromagnetically with a propagation vector $k$ = (1,0,0) and the ordered state moment is 0.20(1) $\mu_B$ along the $b-$axis, in sharp contrast with  the ordered moment of 0.34$-$0.42$\mu_B$ along the $c-$axis observed in CeRu$_2$Al$_{10}$ ($T_N$ = 27 K)~\cite{20},  which is very similar behaviour observed in Ce$_{0.9}$La$_{0.1}$Ru$_2$Al$_{10}$. The $\mu$SR study on Ce(Ru$_{0.97}$Re$_{0.03}$)$_2$Al$_{10}$ reveals the presence of one internal field (frequency) with value of  80 Oe at 1.2 K. The observed single frequency in Ce(Ru$_{0.97}$Re$_{0.03}$)$_2$Al$_{10}$ can be explained by the dipolar field calculation with muon stopping site at 4$c$. On the other hand the $\mu$SR study on Ce$_{1-x}$La$_x$Ru$_2$Al$_{10}$ ($x$ = 0, 0.3 and 0.5) shows the presence of two frequencies.

\par

Further Ce(Os$_{0.97}$Re$_{0.03}$)$_2$Al$_{10}$ has been studied by muon spin relaxation and neutron diffraction measurements. A long-range antiferromagnetic ordering of the Ce sublattice with a substantially reduced value of the magnetic moment 0.18(1) $\mu_B$ along the $c-$axis (same direction as in the undoped system) has been found below $T_N$ = 21 K. On the other hand the electron doping (i.e Ir and Rh) in CeRu$_2$Al$_{10}$ and (Ir) in CeOs$_2$Al$_{10}$ show a large ordered moment of $\approx$1 $\mu_B$ along the $a-$axis. The obtained result reveals the crucial difference between electron- and hole-doping effects on the magnetic ordering in CeT$_2$Al$_{10}$ (T = Ru and Os). The former suppresses the anisotropic $c-f$ hybridization and promotes localized Ce moments controlled by single ion anisotropy. On the contrary, the latter increases the hybridization, keeping the dominant role of the anisotropic exchange on the direction of the moments and shifts the system towards a delocalized nonmagnetic state~\cite{21}.

\par

Finally, it should be pointed out that the obtained results pose a question about the role of the hybridization and the crystal field effects in the reduced moment nature of the magnetic ground state in the undoped CeT$_2$Al$_{10}$ (T = Ru and Os) compounds. The behavior of the system under the hole doping, electron doping and chemical pressure (positive by Y-doping and negative by La-doping) along with the applied  hydrostatic pressure study points to the key role of the hybridization in the anisotropic character of the exchange interactions observed in the undoped compound as well. This also implies the hybridization effect on the moments reduction, attributed by Strigari {\it et. al.}~\cite{36} to the crystal field effects implicitly.

\begin{table}[t]
\begin{center}
\caption{A list of samples studied Ce$_{1-x}$La$_{x}$T$_2$Al$_{10}$ (T = Ru, Os) for $x$ = 0 , 0.1 and their transitions temperatures ($T_N$ = antiferromagnetic ordering temperature), ground state magnetic moment ($\mu_{AF}$) value and moment directions ($\mu_{d}$) obtained from neutron diffraction study.($^*$this work)}
\begin{tabular}{lccccccccccccccc}
\hline
Compounds && &&  $T_N$ &&  $\mu_{AF}$ && $\mu_{d}$ & Spin Gap\\ 
			 && &&         (K) &&  ($\mu_B$) && &(meV)\\ 
\hline
CeRu$_2$Al$_{10}$~\cite{2}   						 &&  && 27.0 &&   0.34(2) && $c-$axis~\cite{2}   &8.0\\ 
Ce$_{0.9}$La$_{0.1}$Ru$_2$Al$_{10}$   &&  && 23.0 &&   0.18(2) &&  $b-$axis$^*$& 7.0\\  
CeOs$_2$Al$_{10}$   						&&  && 28.5   && 0.29(2) &&	$c-$axis~\cite{2} & 11\\  
Ce$_{0.9}$La$_{0.1}$Os$_2$Al$_{10}$  && && 21.7 &&0.23(1) && $c-$axis$^*$ &10\\  
\hline
\end{tabular}
\end{center}
\end{table}

\section{Conclusions} 

We have carried out a comprehensive study on Ce$_{0.9}$La$_{0.1}$T$_2$Al$_{10}$ (T = Ru and Os) using the complementary techniques of magnetization, resistivity, heat capacity, neutron diffraction, inelastic neutron scattering and muon spin relaxation measurements to understand the unusual behavior of the magnetic moment direction and the opening of a spin gap below $T_N$. The neutron diffraction study is unambiguous in confirming the long-range magnetic order in this compound. More interestingly our ND study shows a very small ordered moment of 0.18 $\mu_B$ along the $b-$axis in Ce$_{0.9}$La$_{0.1}$Ru$_2$Al$_{10}$, but a moment of 0.23 $\mu$B along the $c-$axis in Ce$_{0.9}$La$_{0.1}$Os$_2$Al$_{10}$. This contrasting behavior can be explained based on a different degree of hybridization in CeRu$_2$Al$_{10}$ and CeOs$_2$Al$_{10}$: hybridization is stronger in the latter than in the former. Our INS study reveals the presence of a spin gap of 7 meV and 10 meV in the 10\% La-doped T = Ru and Os compounds, respectively. Interestingly the intensity of the spin gap decreases dramatically in La-doped CeOs$_2$Al$_{10}$ compound. 

\par

Further we also present muon spin rotation study on Ce$_{1-x}$La$_x$Ru$_2$Al$_{10}$ ($x$ = 0, 0.3, 0.5 and 0.7), which reveals the presence of two coherent frequency oscillations indicating a long$-$range magnetic ground state in $x$ = 0 to 0.5, but  almost temperature independent Kubo-Toyabe response between 45 mK and 4 K for $x$ = 0.7. The absence of temperature dependent relaxation rate in $x$ = 0.7 despite of the logarithmic rise in the heat capacity down to mK is an unusual behavior. One would expect that near a quantum phase transition, $\mu$SR will sense the presence of quantum fluctuations even in the paramagnetic NFL state. These $\mu$SR results are very similar to that observed in YFe$_2$Al$_{10}$, where despite of NFL behavior observed in the magnetic susceptibility and heat capacity, the $\mu$SR spectra~\cite{2} are independent of temperature down to mK. A major achievement of this work has been the finding of frequency oscillations in our $\mu$SR spectra up to $x$ = 0.5, which for the first time establishes the onset of long-range magnetic ordering in the La-doped compounds and the robust magnetic ordering in spite of the large and anisotropic $c-f$ hybridization in this system. The temperature dependence of the $\mu$SR frequencies and muon depolarization rates of $x$ = 0 follow an unusual behavior with further cooling of the sample below 18 K, pointing at the possibility of another phase transition below 15 K. On the other hand the temperature dependence of the $\mu$SR frequencies and muon depolarization rates of $x$ = 0.3 follows conventional behavior expected for a second order phase transition in the mean field theory.

\section*{ACKNOWLEDGEMENT}

Some of us DTA/ADH would like to thank CMPC-STFC for financial support. The work at Hiroshima University was supported by KAKENHI (Grant No. 26400363) from JSPS, Japan. AMS thanks the SA-NRF (Grant 78832) and UJ Research Committee for financial support. A.B would like to acknowledge FRC of UJ, NRF of South Africa and ISIS-STFC for funding support. We would like to thank P. Manuel, J. M. Mignot and I. Watanabe for an interesting discussion.

\end{document}